\documentstyle[12pt]{article}

\begin{document}
\begin{center}
{\Large \bf The effective inertial acceleration due to
oscillations of the gravitational potential: footprints in the
solar system}

\bigskip

{\large D.L.~Khokhlov}
\smallskip

{\it Sumy State University, R.-Korsakov St. 2, \\
Sumy 40007, Ukraine\\
E-mail: khokhlov@cafe.sumy.ua}
\end{center}

\begin{abstract}
The conjecture is considered that every body induces the wave
field which imposes oscillations on the gravitational potential of
a body. The function for oscillations is chosen to prevent the
gravitational collapse of the matter at the nucleus energy
density. Oscillations of the gravitational potential of a
body produce effective inertial outward acceleration for a
particle orbiting around the body. Footprints of the effective
inertial acceleration due to oscillations of the gravitational
potentials of the Sun and Earth are investigated. The conjecture
allows to explain the anomalous shift of the perihelion of Mercury
and Icarus, the anomalous shift of the perigee of LAGEOS II, the
anomalous acceleration acting on Pioneer 10, 11, the anomalous
increase in the lunar semi-major axis. The advance of the
Keplerian orbit for Earth, Jupiter, Neptune, Uranus caused by the
effective inertial acceleration due to oscillations of the
gravitational potential of the Sun is in agreement with the
observational bounds from the planetary ephemeris.

\end{abstract}

The theory of gravity~\cite{mtw} faces the problem of
singularities. In particular, singularities may arise as a result
of the gravitational collapse of massive stars. A body contracted
to the nucleus energy density resembles the neutron
star~\cite{Sha}. There is a maximum mass for the neutron star of
order of the mass of the Sun, $m_{max}\sim m_{\odot}$. For the
neutron star with the mass less than the maximum mass, the
pressure due to the degenerated neutron Fermi gas balances the
gravity of the star. For the neutron star with the mass more than
the maximum mass, the gravity of the star overcomes the pressure
due to the degenerated neutron Fermi gas, and the star goes to the
singularity.

The radius of the first Bohr's orbit of the Hydrogen atom $a_0$
and the classical radius of electron $r_e$ are related as
\begin{equation}
\frac{r_e}{a_0}=\alpha^2
\label{eq:ra}
\end{equation}
where $\alpha$ is the fine structure constant.
The velocity of the electron at the first Bohr's orbit is given by
\begin{equation}
v_e=\alpha c
\label{eq:ve}
\end{equation}
where $c$ is the velocity of light. The radius of the
nucleus of the Hydrogen atom $r_n$ is of order of the classical
radius of electron
\begin{equation}
r_n\sim r_e.
\label{eq:rr}
\end{equation}
If to contract the Hydrogen atom to the radius
of the nucleus, $r_H\rightarrow r_n$, the fine structure constant
tends to unity, $\alpha\rightarrow 1$, and the velocity of the
electron tends to the velocity of light, $v_e\rightarrow c$. Since
the velocity of light is a limiting one, it is reasonable to think
that the radius of the nucleus is a limiting one, and hence the
nucleus energy density is a limiting one.

It is reasonable to assume that there is some mechanism which
allows to switch off the gravity at a limiting scale thus
preventing appearance of singularities in the gravitational
collapse of the matter. Below we shall consider such a mechanism,
with the nucleus energy density being chosen as a limiting scale.

Assume that every body induces the wave field which
carries scalar perturbation of the gravitational potential.
Thus the wave field imposes oscillations on the
gravitational potential of a body
\begin{equation}
\Phi=-\frac{Gm}{r} + \Delta\Phi\sin\omega_{gr} t
\label{eq:phi}
\end{equation}
where $G$ is the Newtonian constant, $m$ is the mass of the body,
$\omega_{gr}$ is the frequency of oscillations.
On the average oscillations do not change the gravitational
potential of a body.
When a particle orbits around a body, oscillations of the
gravitational potential of the body produce radial
oscillations around the Keplerian orbit of a particle.

Assume that the value of oscillations of the gravitational
potential is a function of the matter density such that the value
of oscillations of the gravitational potential is equal to the
Newtonian potential at the radius of neutron star
\begin{equation}
\Delta\Phi=\frac{Gmr_{NS}^2}{r_{0}^3}
\label{eq:dphi0}
\end{equation}
where $r_0$ is the radius of a body, $r_{NS}$ is its radius of
neutron star.
Suppose that the frequency of oscillations is given by
\begin{equation}
\omega_{gr}=\left(\frac{Gm}{4r_{NS}^3}\right)^{1/2}.
\label{eq:om}
\end{equation}
The frequency of oscillations is large,
$\omega_{gr}\gg 1/T=(Gm/r^3)^{1/2}$.
Then the radial displacement around the Keplerian orbit of a
particle is small~\cite{Lan}
\begin{equation}
\delta r=\frac{\Delta\Phi}{\omega_{gr}^2r}.
\label{eq:dr}
\end{equation}
Kinetic energy due to oscillation motion is given by
\begin{equation}
E_{osc}=\frac{\Delta\Phi^2}{4\omega_{gr}^2r^2}=
\frac{Gmr_{NS}^7}{r_{0}^6r^2}.
\label{eq:E}
\end{equation}
Thus we arrive at the effective Newtonian gravity
\begin{equation}
\Phi=\Phi_N + E_{osc}=\Phi_N
\left(1-\frac{r_{NS}^7}{r_{0}^6r}\right).
\label{eq:mod}
\end{equation}

The value of oscillations of the gravitational potential grows
under the contraction of a body. When the radius of a body reaches
the radius of neutron star, $r_0\rightarrow r_{NS}$, oscillations
of the gravitational potential tends to the value,
$\Delta\Phi\rightarrow Gm/r_{NS}$. The radial displacement due to
oscillations of the gravitational potential tends to the radius of
neutron star, $\delta r\rightarrow r_{NS}$, and the kinetic energy
due to oscillations of the gravitational potential tends to the
value, $E_{osc}\rightarrow Gm/r_{NS}$.
Then oscillations of the gravitational potential switches off the
gravity of a body contracted to the radius of neutron star.
Thus the radius of neutron star is a limiting one. Oscillations of
the gravitational potential prevent the gravitational collapse of
the matter to singularity. Thus the conjecture proposed may
resolve the problem of singularities in the theory of gravity.

Let a particle orbit around a body.
Oscillations of the gravitational potential of the body produce
effective oscillations of the circular velocity of the particle
\begin{equation}
\Delta v=(\Delta\Phi\sin\omega_{gr} t)^{1/2}.
\label{eq:v}
\end{equation}
As a result effective inertial acceleration directed outward the
body arises
\begin{equation}
w_{eff}=\frac{\Delta\Phi}{r}.
\label{eq:wef}
\end{equation}
The effective inertial acceleration does not
modify the Newtonian gravity. Therefore displacement of the radius
of the Keplerian orbit of a particle due to the effective inertial
acceleration cannot be detected in the measurement of the orbit's
radius.

The effective inertial acceleration due to oscillations of the
gravitational potential of a body should lead to the precession of
the Keplerian orbit of a particle under its motion in the
gravitational field of a body. A small addition $\delta \Phi$ to
the potential of a body causes the shift of the perihelion of
particle's orbit per revolution by the value~\cite{Lan}
\begin{equation}
\delta\varphi=\frac{\partial}{\partial M}\frac{2m^{\prime\> 2}}
{M}\int\limits_{0}^{\pi}r^2 \delta\Phi d\varphi
\label{eq:dphi}
\end{equation}
where $m^{\prime}$ is the mass of the particle, $M$ is the angular
momentum of the particle. The non-perturbed orbit is given by
\begin{equation}
r=\frac{p}{1+e\cos\varphi}
\label{eq:orb}
\end{equation}
with
\begin{equation}
p=\frac{M^2}{Gm^{\prime\> 2}m} \qquad
p=a(1-e^2)
\label{eq:p}
\end{equation}
where $p$ is the orbit's latus rectum, $e$ is the eccentricity,
$a$ is the semi-major axis. Let the effective potential due to
oscillations of the gravitational potential of a body
give a small addition to the potential of
a body, $\delta\Phi=\Delta\Phi$. Then integration of
eq.~(\ref{eq:dphi}) with the use of eqs.~(\ref{eq:orb},\ref{eq:p})
gives the shift (advance) of the perihelion of a particle due to
oscillations of the gravitational potential of a body per
revolution
\begin{equation}
\delta\varphi\approx\frac{6\pi a(1-e^2)\Delta\Phi}{Gm}.
\label{eq:dphi2}
\end{equation}
Here we omit the terms containing $\cos^2\varphi$ since an
addition $\delta\Phi$ leads to the deviation from the Keplerian
orbit in this order.

Investigate footprints of the effective inertial acceleration due
to oscillations of the gravitational potential in the solar system.
Suppose that the Sun induces the wave field which imposes
oscillations on the gravitational potential of the Sun. Estimate
the value of oscillations of the gravitational potential of the
Sun adopting the density of neutron star,
$\rho_{NS}\sim 10^{14}\ \mathrm{g/cm^3}$.
Then the radius of neutron star for the Sun is equal to
$r_{NS}=(3m_{\odot}/4\pi\rho_{NS})^{1/3}=
1.7\times 10^{6}\ \mathrm{cm}$.
The value of oscillations of the gravitational
potential of the Sun is equal to
$\Delta\Phi_{\odot}=Gm_{\odot}r_{NS}^2/r_{\odot}^3=
1.1\times 10^{6}\ \mathrm{cm^2/s^2}$.

Observations~\cite{Wei} give the value, $43$ arcseconds per
century, for the anomalous shift (advance) of the perihelion of
Mercury. This anomalous shift is now explained within the
framework of the Einstein general relativity~\cite{Wei}. Suppose
that the value, $43$ arcseconds, is brought about by the effective
inertial acceleration due to oscillations of the gravitational
potential of the Sun. Determine the value of oscillations of the
gravitational potential of the Sun from the
data on the anomalous shift of the perihelion of Mercury with the
use of the Mercury data, the semi-major axis $a=0.387$ AU, the
eccentricity $e=0.206$, the period of revolution $T=0.241$ yr.
Then from eq.~(\ref{eq:dphi2}) we obtain the value of
oscillations of the gravitational potential of the Sun,
$\Delta\Phi_{\odot}=6.4\times 10^5\ \mathrm{cm^2/s^2}$.

Determine the shift of the perihelion of the asteroid Icarus due
to oscillations of the gravitational potential of the Sun with
the use of the Icarus data, the semi-major axis $a=1.076$ AU, the
eccentricity $e=0.827$, the period of revolution $T=1.116$ yr.
In accordance with eq.~(\ref{eq:dphi2}) the value of oscillation
of the gravitational potential of the Sun,
$\Delta\Phi_{\odot}=6.4\times 10^{5}\ \mathrm{cm^2/s^2}$,
gives the shift (advance) of the perihelion of Icarus, $8.5$
arcseconds per century. The observed anomalous shift (advance) is
$9.8\pm 0.8$ arcseconds per century~\cite{Wei}. Thus the
calculated value lies below the uncertainty margin by
around $5\%$ of the observed effect.

Oscillations of the gravitational potential of the Sun yield the
effective outward acceleration of the Earth. This effective
outward acceleration may be seen in ranging of distant
spacecraft. The effective outward acceleration due to oscillations
of the gravitational potential of the Sun should give the blue
shift into the frequency of light at the Earth
\begin{equation}
\frac{\Delta\omega}{\omega}=\frac{\Delta\Phi_{\odot}t}{cr_{SE}}
\label{eq:dom}
\end{equation}
where $r_{SE}$ is the distance between the Earth and Sun,
$t$ is the time of two-leg light travel.
In ranging of distant spacecraft, the acceleration
of the Earth outward the Sun looks like the acceleration of the
spacecraft inward the Sun. Then one may interpret the shift of the
reference frequency given by eq.~(\ref{eq:dom}) as a shift of the
observed re-transmitted frequency due to the acceleration of the
spacecraft inward the Sun. The acceleration of the
Earth gives contribution into the shift of the reference frequency
during the time of two-leg light travel, while the acceleration of
the spacecraft gives contribution into the shift of the observed
re-transmitted frequency during the time of one-leg light travel.
Then, if to interpret the observed anomalous shift of frequency as
the anomalous acceleration of the spacecraft inward the Sun, its
value is doubled in comparison with the effective outward
acceleration of the Earth. Thus the effective outward acceleration
of the Earth due to oscillations of the gravitational potential of
the Sun, $\Delta\Phi_{\odot}=6.4\times 10^{5}\ \mathrm{cm^2/s^2}$,
may be interpreted as the inward acceleration of the spacecraft,
$w=2\Delta\Phi_{\odot}/r_{SE}=
8.5 \times 10^{-8}\ \mathrm{cm/s^2}$.
Analyses of radio Doppler and ranging data from distant spacecraft
in the solar system indicated that an apparent anomalous
acceleration is acting on Pioneer 10 and 11, with a magnitude
$ w_P = (8.74 \pm 1.25) \times 10^{-8}\ \mathrm{cm/s^2}$, directed
towards the Sun~\cite{An}.
Thus the effective outward acceleration of the Earth due to
oscillations of the gravitational potential of the Sun mimics the
anomalous acceleration acting on Pioneer 10 and 11. Thus the value
of oscillations of the gravitational potential of the Sun
explaining the anomalous shift of the perihelion of Mercury may
explain the anomalous acceleration acting on Pioneer 10 and 11.

Oscillations of the gravitational potential of the Sun yield the
radial displacement around the Keplerian orbit of a planet.
In view of eq.~(\ref{eq:dr}), the radial displacement around the
Keplerian orbit of the Earth is of order
$\delta r\sim 10^{-15}\ \mathrm{cm}$.
This value is too small to be seen in observations. The Viking
ranging data~\cite{Vik} determine the Earth-Mars radius
to around a 7.5 m accuracy.

The effective inertial acceleration due to oscillations of the
gravitational potential of the Sun yields the advance of the
Keplerian orbit,
$\delta\varphi\sim 6\pi\Delta\Phi_{\odot}/\Phi_N$, where $\Phi_N$
is the Newtonian potential of the Sun at the radius $r$.
The dark matter restricted within the radius $r$ yields the
advance of the Keplerian orbit,
$\delta\varphi\sim 2\pi \Phi_{DM}/\Phi_N$, where $\Phi_{DM}$ is
the potential of the dark matter restricted within the radius $r$.
Hence the effective inertial acceleration due to oscillations of
the gravitational potential of the Sun mimics the effective mass
restricted within the radius $r$
\begin{equation}
m_{eff}=\frac{\Delta\Phi_{\odot}r}{3G}.
\label{eq:mef}
\end{equation}
Compare the effective mass due to oscillations of the
gravitational potential of the Sun,
$\Delta\Phi_{\odot}=6.4\times 10^{5}\ \mathrm{cm^2/s^2}$,
with the observational bounds for the dark matter from the
planetary ephemeris.
Within the Earth's orbit the effective mass is
$m_{eff}=2.4 \times 10^{-8}m_{\odot}$ while the
observational bound is
$m_{DM}\leq 7 \times 10^{-8}m_{\odot}$~\cite{Mik}.
Within Jupiter's orbit the effective mass is
$m_{eff}=1.2 \times 10^{-7}m_{\odot}$ while the observational
bound is
$m_{DM}=(0.12\pm 0.027) \times 10^{-6}m_{\odot}$~\cite{An2}.
Within Uranus's orbit the effective mass is
$m_{eff}=4.7 \times 10^{-7}m_{\odot}$ while the observational
bound is $m_{DM}\leq 0.5 \times 10^{-6}m_{\odot}$~\cite{An2}.
Within Neptune's orbit the effective mass is
$m_{eff}=7.3 \times 10^{-7}m_{\odot}$ while the
observational bound is
$m_{DM}\leq 3 \times 10^{-6}m_{\odot}$~\cite{An2}.
Thus the effective mass due to oscillations of the
gravitational potential of the Sun is in agreement
with the observational bounds for the Earth's,
Jupiter's, Neptune's, and Uranus's orbits.

Oscillations of the gravitational potential of the Sun yield the
effective inertial acceleration of the Moon with respect to the
Earth, $w\approx\Delta\Phi_{\odot}r_{EM}\sin\Omega t/r_{SE}^2$,
where $r_{EM}$ is the distance between the Earth and Moon,
$\Omega$ is the angular velocity of the Moon. This leads to the
polarization of the Moon's orbit in the direction of the Sun, with
the Earth-Moon distance decreasing. As a result the Moon's
circular velocity increases by the value,
$\Delta v=(\sqrt{2}\Delta\Phi_{\odot}r_{EM}/2r_{SE})^{1/2}$,
average for the period of revolution of the Moon. Then inertial
acceleration of the Moon outwards the Earth increases by the
value, $w=\sqrt{2}\Delta\Phi_{\odot}/2r_{SE}$,
average for the period of revolution of the Moon.
Oscillations of the gravitational potential of the Sun,
$\Delta\Phi_{\odot}=6.4\times 10^{5}\ \mathrm{cm^2/s^2}$,
yield the effective inertial acceleration of the Moon
outwards the Earth, $w=3.0 \times 10^{-8}\ \mathrm{cm/s^2}$.

As follows from telescopic observations of the
Earth's rotation collected for the last 300 years,
e.g.~\cite{Per}, the rate of deceleration of the Earth's rotation
is $\dot{T}_E=1.4\times 10^{-5}\ \mathrm{s/yr}$. As a result
the lunar semi-major axis increases with a rate,
$\dot{a}_{tel}=2.53\ \mathrm{cm/yr}$. Measurement of the
Earth-Moon distance in lunar laser ranging~\cite{Dic} gives the
value, $\dot{a}_{LLR}=3.82\ \mathrm{cm/yr}$, hence points to the
anomalous increase in the lunar semi-major axis with an excessive
rate, $\dot{a}_{LLR}-\dot{a}_{tel}=1.29\pm 0.2\ \mathrm{cm/yr}$.
This corresponds to the anomalous outward acceleration of the Moon,
$w=(\dot{a}_{LLR}-\dot{a}_{tel})c/r_{EM}
=(3.2\pm 0.5)\times 10^{-8}\ \mathrm{cm/s^2}$.
Thus the effective inertial acceleration of the Moon due to
oscillations of the gravitational potential of the Sun explains
the main part (around $94\%$) of the anomalous outward
acceleration of the Moon.

Suppose that the Earth induces the wave field which imposes
oscillations on the gravitational potential of the Earth. Estimate
the value of oscillations of the gravitational potential of the
Earth. The radius of neutron star for the Earth is equal to
$r_{NS}=(3m_{\oplus}/4\pi\rho_{NS})^{1/3}=
2.4\times 10^{4}\ \mathrm{cm}$.
The value of oscillations of the gravitational potential
of the Earth is equal to
$\Delta\Phi_{\oplus}=Gm_{\oplus}r_{NS}^2/r_{\oplus}^3=
9\times 10^{2}\ \mathrm{cm^2/s^2}$.

Observations of the Earth's satellite LAGEOS II~\cite{Ior} give
the value, $3.4$ arcseconds per year,
for the anomalous shift (advance) of the perigee of LAGEOS II.
This anomalous shift is now explained within the framework of the
Einstein general relativity~\cite{Ior}. Suppose that the value,
$3.4$ arcseconds, is brought about by the effective inertial
acceleration due to oscillations of the gravitational potential of
the Earth. Determine the value of
oscillations of the gravitational potential of the Earth
from the data on the anomalous shift of the perigee of LAGEOS II
with the use of the LAGEOS II data, the semi-major axis
$a=1.2163\times 10^{9}\ \mathrm{cm}$, the eccentricity $e=0.014$,
the period of revolution $T=3.758$ hr, the inclination
$I=52.65^0$. Then from eq.~(\ref{eq:dphi2}) we obtain the
value of oscillations of the gravitational potential of the Earth,
$\Delta\Phi_{\oplus}=1.2\times 10^2\ \mathrm{cm^2/s^2}$.
Estimate the effect of oscillations of the gravitational potential
of the Sun on the motion of LAGEOS II,
$\Delta\Phi=\sqrt{2}\Delta\Phi_{\odot}a\cos I/2r_{SE}=
0.3\times 10^2\ \mathrm{cm^2/s^2}$.
In view of this effect, we obtain the
value of oscillations of the gravitational potential of the Earth,
$\Delta\Phi_{\oplus}=0.9\times 10^2\ \mathrm{cm^2/s^2}$.

Oscillations of the gravitational potential of the Earth,
$\Delta\Phi_{\oplus}=0.9\times 10^2\ \mathrm{cm^2/s^2}$,
yield the effective inertial outward acceleration of the Moon,
$w=\Delta\Phi_{\oplus}/r_{EM}=0.2\times 10^{-8}\ \mathrm{cm/s^2}$.
This acceleration explains around $6\%$ of the anomalous outward
acceleration of the Moon. The overall effective inertial
acceleration of the Moon due to oscillations of the gravitational
potential of the Sun and Earth is
$w=3.0+0.2=3.2\times 10^{-8}\ \mathrm{cm/s^2}$ and thus completely
explains the anomalous outward acceleration of the Moon.

We have considered the conjecture that every body induces the wave
field which imposes oscillations on the gravitational potential of
a body. The function for oscillations is chosen to prevent the
gravitational collapse of the matter at the nucleus energy
density. The radial displacement around the Keplerian orbit of a
planet due to oscillations of the gravitational potential of the
Sun is too small to be seen in observations. Oscillations of the
gravitational potential produce
effective inertial outward acceleration for a particle orbiting
around the body. Footprints of the effective inertial acceleration
due to oscillations of the gravitational potential of the Sun may
be revealed as an anomalous shift of the perihelion of the
Keplerian orbit of a planet or as an anomalous shift of the
frequency of light seen in ranging. Although we have rough
estimation of the value of oscillations of the gravitational
potential of the Sun
it is possible to perform the accurate testing of the conjecture
by comparing the results of different observations. The value of
oscillations of the gravitational potential of the Sun obtained
from the anomalous shift of the perihelion of Mercury is
consistent with that obtained from the anomalous shift of the
frequency of light at the Earth now interpreted as the anomalous
acceleration acting on Pioneer 10, 11. Also the value of
oscillations of the gravitational potential of the Sun
approximately explains the anomalous shift (advance) of the
perihelion of Icarus, with the calculated value lies below the
uncertainty margin by around $5\%$ of the observed effect. Also
the value of oscillations of the gravitational potential of the
Sun explains the main part (around $94\%$) of the anomalous
outward acceleration of the Moon. The value of oscillations of the
gravitational potential of the Earth obtained from the anomalous
shift of the perigee of LAGEOS II explains the retaining part
(around $6\%$) of the anomalous outward acceleration of the Moon.
The advance of the Keplerian orbit for Earth, Jupiter,
Neptune, Uranus caused by the effective inertial force due to
oscillations of the gravitational potential of the Sun is in
agreement with the observational bounds from the planetary
ephemeris.

The conjecture considered provides an explanation of the anomalous
precession (advance) of the Keplerian orbit alternative to the
general relativity.
So from the point of view of the conjecture considered the
general relativity is questioned. Note that excluded the
anomalous precession (advance) of the Keplerian orbit we may
describe the planetary motion in the solar system within the
Newtonian mechanics. At the same time the relativistic effects for
the electromagnetic waves in the field of the gravitational
potential are confirmed by the experiments. If to adopt the
conjecture considered we should reinterpret the theory of
relativity in the way that relativistic effects pertain only to
the electromagnetism while the gravitation remains
non-relativistic. That is the motion of the electromagnetic waves
follows the relativistic dynamics while the motion of the massive
bodies follows the Newtonian dynamics.

\end{document}